\documentclass[pra,twocolumn,aps,groupedaddress,longbibliography,nofootinbib]{revtex4-1}  
\usepackage{hyperref}
\usepackage{graphicx}
\usepackage{amsmath}
\usepackage{amsfonts}
\usepackage{amssymb}
\usepackage{commath}
 \usepackage{mathtools}
\usepackage{epsfig}
\usepackage{subfigure}
\usepackage{mathtools}
\usepackage{braket}
\usepackage{float}
\usepackage[usenames,dvipsnames]{color}
\usepackage{setspace}
\usepackage{bm}
\usepackage{times}
\hypersetup{
      colorlinks=true,
      citecolor=blue,
      linkcolor=blue,
      urlcolor=blue}

\newcommand{\add}{a_{dd}}
\newcommand{\edd}{\epsilon_{dd}}
\newcommand{\br}{\mathbf{r}}
\newcommand{\bx}{\mathbf{x}}

\newcommand{\LGP}{\mathcal{L}_\text{EGP}}
\newcommand{\gammaQF}{\gamma_{\mathrm{QF}}}

\begin{document}
 \title{Numerical calculation of dipolar quantum droplet stationary states}
\author{Au-Chen Lee}
\author{D.~Baillie} 
\author{P.~B.~Blakie}  
\affiliation{The Dodd-Walls Centre for Photonic and Quantum Technologies, Department of Physics, University of Otago, New Zealand}

\begin{abstract}
We describe and benchmark a method to accurately calculate the quantum droplet states that can be produced from a dipolar Bose-Einstein condensate. Our approach also allows us to consider vortex states, where the atoms circulate around the long-axis of the filament shaped droplet. We apply our approach to determine a phase diagram showing where self-bound droplets are stable against evaporation, and to quantify the energetics related to the fission of a vortex droplet into two non-vortex droplets.
  \end{abstract} 
  
\maketitle
\section{Introduction}
Dipolar condensates consist of highly magnetic atoms  that interact with a long-ranged and anisotropic dipole-dipole interaction (DDI).  
Experiments with dipolar condensates of dysprosium \cite{Kadau2016a,Ferrier-Barbut2016a,Schmitt2016a} and erbium \cite{Chomaz2016a} atoms have prepared the system into one or several self-bound droplets that can preserve their form, even in the absence of any external confinement. These droplets occur in the dipole-dominated regime, where the short ranged $s$-wave interactions are tuned to be weaker than the DDIs  \cite{Baillie2016b,Wachtler2016b}. In this regime the standard meanfield theory provided by the Gross-Pitaevskii equation is inadequate, as it predicts that the condensate is unstable to mechanical collapse. Here (beyond meanfield) quantum fluctuation corrections become important  \cite{LHY1957,Lima2011a,Lima2012a} and stabilize the droplets against collapse \cite{Petrov2015a,Saito2016a,Wachtler2016a,Bisset2016a}.  Droplets have been produced with $10^3$--$10^4$ atoms and  peak densities roughly an order of magnitude higher than that of typical condensates. These droplets are still well within the dilute regime  (i.e.~$na^3\ll1$, where $n$ is the density and $a$ is the interaction length scale). In this regime the extended Gross-Pitaevskii equation (EGPE)  is expected to provide a good description, and indeed the EGPE has been successfully used to model the equilibrium and dynamical properties of droplets.  The excitation spectrum of the droplets has been the subject of theoretical and experimental studies \cite{Wachtler2016b,Chomaz2016a,Baillie2017a}, and more recently the possibility of preparing a droplet in an excited vortex state has been considered \cite{Cidrim2018a,Lee2018a}, although these have been found to be unstable.

So far little attention in the literature has been given to the details of EGPE calculations for stationary dipolar droplet states.
The DDI needs to be treated with care since it is singular and long-ranged. Because the droplets are small, dense,  and highly elongated (taking a filament-like shape), it becomes more difficult to treat the DDI accurately without taking large and dense numerical grids.  Due to these technical challenges few groups outside of those already working on dipolar condensates have reported calculations for these droplets.
 In contrast we note that quantum droplets have also been realized in two component condensates \cite{Petrov2015a,Cabrera2018a,Semeghini2018a,DErrico2019a}. The absence of DDIs in these droplets makes the calculations more straightforward, and a rather large number of theory groups have reported work on this system.

In this paper we aim to outline a numerical technique to calculate self-bound quantum droplet states, including the case where the droplet has a vortex.
We demonstrate the importance of using a truncated interaction potential to accurately evaluate the DDI energy, and introduce a simple gradient flow technique for obtaining the stationary solutions of the EGPE.   We present benchmark results for the energy and chemical potential for vortex and non-vortex droplet stationary states. As applications we calculate a phase diagram for the stability of the droplet states, and quantify the tendency for vortex droplets to undergo fission into a pair of non-vortex droplets.

The outline of the paper is as follows. In Sec.~\ref{Sec:Formalism} we introduce the  EGPE for describing quantum droplet states. We also rewrite the formalism in cylindrical coordinates and introduce our dimensionless units. In Sec.~\ref{SEC:NBCquad} we introduce the Bessel-cosine based method that we use to discretize the EGPE and to enforce states to have a particular value of angular momentum. A variational approach is also presented and used to test the accuracy of the discretization when evaluating the various energy terms related to the EGPE. The gradient flow (or imaginary time) method is introduced in Sec.~\ref{SEC:Gflow} to obtain energy minimising solutions of the EGPE. The main results are presented in Sec.~\ref{SEC:Results} including: benchmark results for droplet energies and chemical potentials, a phase diagram indicating where the droplets are stable to evaporation, and results quantifying the energetic instability of vortex droplets to fission. We then conclude in Sec.~\ref{SEC:Conclusions}.

\section{Formalism}\label{Sec:Formalism}
\subsection{Extended Gross-Pitaevskii Equation}

 \subsubsection{General formulation}
Several works \cite{Saito2016a,Ferrier-Barbut2016a,Wachtler2016a,Schmitt2016a,Wachtler2016b,Bisset2016a,Baillie2016b,Chomaz2016a,Boudjemaa2015a,Boudjemaa2016a,Boudjemaa2017a,Oldziejewski2016a,Macia2016a} have  established that the ground states and dynamics of a dipolar condensate in the droplet regime is well-described by the EGPE. In this formalism the time-independent stationary state wavefunction $\Psi$ is a solution of $\mu\Psi=\LGP[\Psi]$ where 
\begin{align}
    \LGP[\Psi]\! &\equiv\!  \biggl[-\frac{\hbar^2\nabla^2}{2M} \!+\! V_\text{tr} \!+\! g_s|\Psi|^2 \!+\!  g_{dd}\Phi(\bx) \!+\! \gammaQF |\Psi|^3\biggr]\!\Psi.\label{e:LGP}
\end{align}  
Here $V_\text{tr}$ describes any trapping potential, $\mu$ is the chemical potential and  $g_s=4\pi a_s \hbar^2/M$ is the $s$-wave coupling constant, with $a_s$ being the $s$-wave scattering length.
The   potential 
\begin{align}
 \Phi (\bx) =  \int d\bx' \,I(\bx\!-\!\bx')|\Psi(\bx')|^2,
 \end{align}
  describes the effects of the long-ranged DDIs  where the kernel is
\begin{align}
   I(\br) &= \frac{3}{4\pi}\frac{ 1-3\cos^2 \theta}{r^3}.\label{UDDI}
\end{align}
This is written for the case of dipoles polarized along the $z$ axis by an external field, where $\theta$ is the angle between $\br$ and the $z$ axis.
The dipole coupling constant is $g_{dd}=4\pi \add \hbar^2/M$, where $\add=M\mu_0\mu_m^2/12\pi\hbar^2$ is the dipole length determined by the magnetic moment $\mu_m$ of the particles. The leading-order quantum fluctuation correction to the chemical potential for a uniform system of density $n$ is $\Delta \mu = \gammaQF n^{3/2}$,
where $\gammaQF\! =\! \frac{32}{3}g_s\sqrt{\frac{a_s^3}{\pi}}(1+\tfrac32 \edd^2)$ 
and $\edd\equiv\add/a_s$ \cite{Lima2011a,Ferrier-Barbut2016a,Bisset2016a}. The effects of quantum fluctuations are included in Eq.~\eqref{e:LGP} using the local density approximation $n \rightarrow |\Psi(\bx)|^2$.
Stationary EGPE states are also local minima of the energy functional
\begin{align}
E[\Psi]=E_\text{kin}+E_\text{tr}+E_\text{int}+E_\text{QF},\label{Eq3DEfunc}
\end{align}
where 
\begin{align}
E_\text{kin}&=-\frac{\hbar^2}{2M}\int d\mathbf{x}\Psi^*\nabla^2\Psi,\\
E_\text{tr}&= \int d\mathbf{x}\Psi^*V_\text{tr}\Psi,\\
E_\text{int}&= \frac{1}{2}\int d\mathbf{x}\Psi^*\left(g_s|\Psi|^2+g_{dd}\Phi\right)\Psi,\\
E_\text{QF}&= \frac{2}{5}\gamma_\text{QF}\int d\mathbf{x} |\Psi|^{5},
\end{align}
represent the kinetic, trap potential, interaction and quantum fluctuation energies, respectively.

\subsubsection{Dimensionless cylindrical formulation}
We now write the problem in a form utilizing the cylindrical symmetry and introducing natural units.
While we do not present any results here that include a trapping potential (we focus on self-bound states in the absence of confinement), for generality we include a cylindrically symmetric trapping potential in our formulation. We take this to be of the form
$V_{\mathrm{tr}}=\frac{1}{2}M(\omega_\rho^2\rho^2+\omega_z^2z^2)$,
where  $\rho=\sqrt{x^2+y^2}$,  and $\{\omega_\rho,\omega_z\}$ are the trap frequencies. For this case the system is cylindrically symmetric (since we have also chosen the dipoles to be along $z$) and we can choose to look for stationary solutions of the form 
\begin{align}
\Psi(\mathbf{x})=\psi(\rho,z)e^{is\phi},\label{psis}
\end{align}
where  $\psi$ is real, $\phi=\arctan(y/x)$ and $s$ is an integer specifying the $z$-component angular momentum of the state. 
By separating variables, and introducing units of length $x_0=a_{dd}$ and energy $E_0=\hbar^2/Ma_{dd}^2$, we arrive at the effective cylindrical GPE 
\begin{align}
\mu\psi&=\mathcal{L}[\psi],\label{cylGPE}\\
\mathcal{L}[\psi]&\equiv  h_{sp} \psi+4\pi\left({\epsilon_{dd}^{-1}}\psi^2+  \Phi\right) \psi+\gammaQF|\psi|^3\psi.\label{Ls}
\end{align}
Here
\begin{align}
    h_{sp} =-\frac{1}{2}\left(D_s+\frac{\partial^2}{\partial z^2}\right) +\frac{1}{2}({\bar\omega}_\rho^2\rho^2+{\bar\omega}_z^2z^2),\label{hsp}
\end{align}
is the single particle Hamiltonian, which features the Bessel differential operator
\begin{align}
D_s\equiv\frac{\partial^2}{\partial\rho^2}+\frac{1}{\rho}\frac{\partial}{\partial \rho}-\frac{s^2}{\rho^2},\label{BesselOp}
\end{align} 
and $\bar\omega_\nu=\hbar\omega_\nu/E_0$, $\nu=\{\rho,z\}$. 
We also note that for this choice of units the nonlinear coupling constants are $g_s\to4\pi\epsilon_{dd}^{-1}$, $g_{dd}\to4\pi$, and  
\begin{align}
\gammaQF=\frac{4}{3\pi^2}\left(\frac{4\pi}{\epsilon_{dd}}\right)^{5/2} \left(1+\frac{3}{2}\epsilon_{dd}^2\right),
\label{numgQF}
\end{align}
thus all specified in terms of $\epsilon_{dd}$.

The convolution used to evaluate $\Phi $ is performed in three-dimensions, but the resulting $\Phi $ is a cylindrically symmetric function:
\begin{align}
\Phi (\rho,z)&\equiv \int d\mathbf{x}'\,I(\mathbf{x}-\mathbf{x}')|\psi(\rho',z')|^2,\label{PhiDrhoz}\\
&=\int \frac{dk_\rho\,dk_z}{(2\pi)^2}e^{ik_zz}k_\rho J_0(k_\rho\rho)\tilde{I}(k_\rho,k_z)\tilde{n}(k_\rho,k_z),\label{PhiDrhoz1} 
\end{align}
where the Fourier transformed density and DDI kernel are
\begin{align}
\tilde{n}(k_\rho,k_z)&=2\pi \int d\rho\,dz\,\rho J_0(k_\rho\rho)e^{-ik_zz}|\psi(\rho,z)|^2,\label{ntilde0}\\
\tilde{I}(k_\rho,k_z)& =  \frac{3k_z^2}{k_\rho^2+k_z^2}-1 ,\label{Iddk1}
\end{align}
respectively.
We also note that here we choose the condensate to be  normalized to the number of atoms $N$, i.e.~
\begin{align}
 2\pi \int _0^\infty d\rho \int_{-\infty}^{\infty}dz\,\rho |\psi|^2=N.
\end{align}

\section{Bessel-cosine numerical representation}\label{SEC:NBCquad}
To accurately treat the terms appearing  in the EGPE operator we use a different discretization for the radial and axial dimensions, and we discuss these separately in the next subsections.

\subsection{Radial Bessel treatment}\label{SecRadial}
\subsubsection{Bessel grid and quadrature}\label{SecRadialGrid}
In the radial direction we consider $N_\rho$ points, that non-uniformly span the interval $(0,R)$, given by 
\begin{align}
\rho_{qi}=\frac{\alpha_{qi}}{K_q},\qquad i=1,\ldots, N_\rho, \label{rho_grid}
\end{align}
which we refer to as the $q$-order Bessel grid.
Here $K_q= {\alpha_{q\,N_\rho+1}}/R$ , and $\{\alpha_{qi}\}$ are the ordered non-zero roots of the Bessel function $J_q(x)$ of integer order $q$.   In this work we will need several such radial grids, each with the same number of points and range but of different orders. For this reason we need to adopt a notation that explicitly indicates the order.
We also introduce the reciprocal space grid of the same order
\begin{align}
k_{qi}=\frac{\alpha_{qi}}{R},
\qquad i=1,\ldots, N_\rho,\label{krho-grid}
\end{align}
that spans the interval $(0,K_q)$.

The radial grid points can be associated with a quadrature-like integration formula \cite{Lemoine1994a,Ghanem1999a}
\begin{align}
I[g(\rho)]=\int_0^\infty d\rho\,\rho g(\rho)\approx\sum_{i=1}^{N_\rho}w_{qi}\,g_{qi},\label{Quadr}
\end{align}
where $w_{qi}$ are the real-space quadrature weights
\begin{align}
w_{qi}&=\frac{2}{K_q^2|J_{q+1}(\alpha_{qi})|^2},
\end{align}
and we have used the notation $g_{qi}=g(\rho_{qi})$ to denote the function $g(\rho)$ sampled on the $q$-order grid.
This integration requires that the functions of interest have limited spatial range, i.e.~$g(\rho>R)=0$.  

Similarly, for  the reciprocal $k_\rho$-space we have a quadrature-like integration formula for a  function $\tilde{g}(k_\rho)$
\begin{align}
I[\tilde{g}(k_\rho)]=\int_0^\infty d k_\rho\,k_\rho \tilde{g}(k_\rho)\approx\sum_{i=1}^{N_\rho}\tilde{w}_{qi}\,\tilde{g}_{qi} ,\label{Quadk}
\end{align}
where $\tilde{w}_{qi}$ are the $k_\rho$-space quadrature weights
\begin{align}
\tilde{w}_{qi}&=\frac{2}{R^2|J_{q+1}(\alpha_{qi})|^2},
\end{align}
and $\tilde{g}_{qi}=\tilde{g}(k_{qi})$.
Result (\ref{Quadk}) is only valid on our  grid if the function is bandwidth limited, i.e.~$\tilde{g}(k_\rho>K_q)=0$.

\subsubsection{Hankel transformation}
The Bessel grid is useful because it allows an accurate two-dimensional (2D) Fourier transformation of functions of the form 
\begin{align}
F(\bm{\rho})=f(\rho)e^{im\phi},\label{EqFm}
\end{align}
 where we have used $\bm{\rho}$ to denote the planar position vector with polar coordinates ($\rho,\phi)$. The 2D Fourier transform  of $F$ is 
\begin{align}
\tilde{F}(\mathbf{k}_\rho)&= \int d\bm{\rho}\,e^{-i\mathbf{k}_\rho\cdot\bm{\rho}}F(\bm{\rho})
 =2\pi i^{-m}e^{im\phi_k}\tilde{f}(k_\rho),\label{FTradial}
 \end{align}
 where
 \begin{align}
\tilde{f}(k_\rho)= &\int_0^\infty\! d\rho\,\rho J_m(k_\rho\rho)\,f(\rho),\label{Hankelq}
\end{align}
is the  $m$-th order Hankel transform, arising because the angular integral of $e^{im\phi-i\mathbf{k}_\rho\cdot\bm{\rho}}$ yields the $J_m$ Bessel function. Here we have introduced $\mathbf{k}_\rho$ to represent the 2D $k$-space vector, with
polar coordinates $(k_\rho,\phi_k)$. 

For the case where the angular momentum of the function [i.e., $m$ from Eq.~(\ref{EqFm})] and the order of the grid used to sample the function   are the same (i.e.~$q=m$) an accurate discrete Hankel transform can be implemented. For this case the transform yields
$\tilde{f}$ sampled on the $k_{qi}$ grid from $f$ sampled on the $\rho_{qi}$ grid (see  \cite{Guizar-Sicairos04}). The explicit form of this discrete transform is obtained by evaluating (\ref{Hankelq}) using the quadrature formula (\ref{Quadr}) 
\begin{align}
\tilde{f}_{qi}=\sum_j\mathcal{H}_{ij}{f}_{qj},\label{discreteHT}
\end{align}
where 
\begin{align}
\mathcal{H}_{ij}=w_{qj}J_q(k_{qi}\rho_{qj})=\frac{2J_q\left(\frac{\alpha_{qi}\alpha_{qj}}{\alpha_{qN_\rho+1}}\right)}{K_q^2|J_{q+1}(\alpha_{qj})|^2},\label{HankelH}
\end{align}
is the $q$-order Hankel transformation matrix and ${f}_{qj}={f}(\rho_{qj})$ etc.   

Similarly, the inverse 2D transform 
\begin{align}
F(\bm{\rho})=\int d\mathbf{k}_\rho \frac{e^{i\mathbf{k}_\rho\cdot\bm{\rho}}}{(2\pi)^2}\tilde{F}(\mathbf{k}_\rho)= \frac{ i^{-q}e^{iq\phi}}{2\pi}{f}(\rho),\label{iFTradial}
\end{align} 
is accomplished by the inverse Hankel transform
\begin{align}
f (\rho)=\int_0^\infty dk_\rho\,k_\rho J_q(k_\rho\rho)\tilde{f} (k_\rho),
\end{align}
with discrete form ${f}_{qi}=\sum_j\mathcal{H}_{ij}^{-1}\tilde{f}_{qj}$, with 
$
\mathcal{H}_{ij}^{-1}=\frac{K_q^2}{R^2}\mathcal{H}_{ij}$.
The Discrete Hankel transform matrices are not exact inverses of each other, but for typical grid sizes ($N_\rho\sim10^2$) we have that $\sum_j\mathcal{H}_{ij} \mathcal{H}_{jk}^{-1}\approx\delta_{ik}+O(10^{-9})$, which is adequate for our purposes.

\subsubsection{Radial Laplacian operator}
Hankel transforms are particularly useful for accurately evaluating the kinetic energy operator in radially symmetric cases (e.g.~see \cite{Bisseling1985}). To see this we note that by separating variable the 2D Laplacian $\nabla^2_{\bm{\rho}}$ acting on the function  $f(\rho)e^{is\phi}$ is equivalent to the Bessel differential operator $D_s$  (\ref{BesselOp}) acting on $f (\rho)$ [cf.~Eq.~(\ref{hsp})]. Using that
 $J_s$ are eigenfunctions of  $D_s$, i.e.~$D_s J_s(k_\rho\rho)=-k_\rho^2J_s(k_\rho\rho)$, we can utilize the Hankel transform to act on a radial function with $D_s$: 
 \begin{align}
[D_s f]_i\approx-\sum_{jk}\mathcal{H}^{-1}_{ij}{k_{sj}^2}\mathcal{H}_{jk}f_{k}.\label{DHTlaplacian}
 \end{align}
 Allowing us to implement a spectrally accurate radial kinetic energy matrix [i.e.~discretized version of $T_\rho=-\frac{1}{2}D_s$, c.f.~Eq.~(\ref{hsp})]:
 \begin{align}
T_{\rho,ij} =\frac{1}{2}\sum_{jk}\mathcal{H}^{-1}_{il}k_{sl}^2\mathcal{H}_{lj}.\label{THankel}
\end{align}

\subsection{Axial cosine treatment}\label{SecAxial}
\subsubsection{Trigonometric grids and quadrature}\label{SecAxialGrid}

Our system has reflection symmetry along $z$ so functions of interest will be of definite parity and here we will concern ourselves with the case of even parity.  Utilizing this symmetry we use a half-grid on the interval $(0,Z)$ spanned by  $N_z$ equally spaced points  \begin{align}
z_j=(j-\tfrac{1}{2})\Delta z,\qquad j=1,2,\ldots N_z,
\end{align} 
where $\Delta z=Z/N_z$. The corresponding reciprocal grid 
\begin{align}
k_{j}=(j-\tfrac{1}{2})\Delta k,\qquad j=1,2,\ldots N_z,
\end{align} 
spans the interval $(0,K_z)$ where $\Delta k=\pi/Z$ and $K_z=\pi/\Delta z$.
The appropriate quadrature for this grid is the rectangular rule
\begin{align}
I[g(z)]=\int_{-\infty}^{\infty}dz\,g(z)\approx\sum_{i=1}^{N_z} g_i\,2\Delta z,
\end{align}
where $g_i=g(z_i)$.
 
\subsubsection{Cosine transformations}
The 1D Fourier transform for an even parity function is given by the cosine transform
\begin{align}
\tilde{f}(k_z)&=\int_{-\infty}^{\infty}dz\, e^{-ik_zz}{f}(z),\label{FTaxial}\\
 &=2\int_{0}^{\infty}dz \cos(k_zz){f}(z).  \label{costransf}
\end{align}
We can discretize this transform onto our chosen grid as
\begin{align}
\tilde{f}_i&=\Lambda_{ij}f_j,\label{DSCT}
\end{align}
where
\begin{align}
{\Lambda}_{ij}=
2\cos(k_{i}z_j)\Delta z,  
\end{align}
is the transformation matrix,   $f_j=f(z_j)$, and $\tilde{f}_i=\tilde{f}(k_i)$. This can be identified as the type-IV discrete cosine transformation (e.g.~see \cite{Strang1999a}).

The inverse Fourier transformation  
${f}(z)=\frac{1}{2\pi}\int_{-\infty}^{\infty}dk_z e^{ik_zz}\tilde{f}(k_z)$ 
similarly can be mapped to the inverse discrete cosine transform
$f_j=\Lambda^{-1}_{ij}\tilde{f}_j$,
where  
${\Lambda}_{ij}^{-1}= \frac{K_z}{2\pi Z}{\Lambda}_{ij}$,
with $\sum_j{\Lambda}_{ij}^{-1}{\Lambda}_{jk}=\delta_{ik}$.
 
\subsubsection{Axial Laplacian operator}
Utilizing that the derivative operator is diagonal in $k_z$-space, we can implement a discrete second derivative operator as
\begin{align}
\left[\frac{d^2f}{dz^2}\right]_i&\approx
-\sum_{jk}\Lambda_{ij}^{-1}k_{j}^2{\Lambda}_{jk}f_k 
\end{align}
This allows us to define a spectrally accurate axial kinetic energy matrix [i.e.~discretized version of $T_z=-\frac{1}{2}\frac{d^2}{dz^2}$, c.f.~Eq.~(\ref{hsp})]:
\begin{align}
T_{z,ij}= 
\frac{1}{2}\sum_{jk}\Lambda_{ij}^{-1}k_{l}^2{\Lambda}_{lj}.\label{Tcos}\end{align}

\subsection{Treatment of EGPE operators} \label{Sec:EGPEOps}
To find eigenstates of the effective cylindrical EGPE problem (\ref{cylGPE})  we combine the radial and axial treatments described above to define a cylindrical (2D) mesh of points $\bm{\rho}_{ij}\equiv(\rho_{si},z_j)$, choosing the radial order $q$ to match the stationary state circulation $s$.  
 
\subsubsection{Single particle operators}
The single particle operator (\ref{hsp})
acting on the discretized field  $\psi_{ij}=\psi(\bm{\rho}_{ij})$ can be evaluated as
\begin{align}
h_{ij}[\psi]=\sum_kT_{\rho,ik}\psi_{kj} +\sum_k  T_{z,jk}\psi_{ik}+V_{\mathrm{tr},ij}\psi_{ij},
\end{align}
where  
$V_{\mathrm{tr},ij}=\frac{1}{2}(\bar{\omega}_\rho^2\rho_{si}^2+\bar{\omega}_z^2 z_j^2)$,
and we have used the Bessel (\ref{THankel}) and cosine (\ref{Tcos}) derivative operators.

\subsubsection{Local interaction terms}
 The contact interaction term in Eq.~(\ref{Ls}) is a nonlinear term that is local in position space and is evaluated as
 \begin{align}
C_{ij}\left[\psi\right] \equiv4\pi\epsilon_{dd}^{-1} \psi_{ij}^3 ,\label{C}
 \end{align}
 and similarly for the quantum fluctuation term   \begin{align}
Q_{ij}\left[\psi\right] \equiv\gammaQF\left|\psi_{ij}\right|^3 \psi_{ij}.\label{Q}
 \end{align}
 Note: we have assumed that $\psi_{ij}$ is real, but modulus sign is needed on the quantum fluctuation term to properly deal with any case where $\psi_{ij}$ is negative.

\subsubsection{DDI term}\label{DDIterm}
 The DDI potential $\Phi$ can be evaluated using the convolution theorem in cylindrical coordinates, as given in Eq.~(\ref{PhiDrhoz1}).
 To evaluate this expression we first need to Fourier transform the density (\ref{ntilde0}), which we obtain by applying the cosine transform along $z$ and the quadrature rule to evaluate the radial transform  
 \begin{align}
 \tilde{n}_{ij}^\prime=2\pi \sum_{kl}\Lambda_{jl}w_{sk}J_0(k_{0i}\rho_{sk})\psi_{kl}^2.\label{ntilde}
 \end{align}
 Here we use the prime to indicate that the Fourier transformed density is evaluated on the cylindrical $k$-space mesh of grid points  $\mathbf{k}_{ij}'=(k_{0i},k_j)$, noting that the radial $k$-space points are defined for the 0-order Bessel grid\footnote{The radial transform in Eq.~(\ref{ntilde}) is identical to (\ref{discreteHT})  if $s=0$.}.  We then evaluate the potential $\Phi$ from Eq.~(\ref{PhiDrhoz1}) as
  \begin{align}
\Phi_{ij}&=\frac{1}{2\pi}\sum_{kl}\Lambda^{-1}_{jl}\tilde{w}_{0k}J_0(k_{0k}\rho_{si})\tilde{I}(\mathbf{k}_{kl}^\prime)\tilde{n}^\prime_{kl},\label{PhiDDij}
 \end{align} 
and thus the full DDI term  
 \begin{align}
{D}_{ij}\left[\psi\right]\equiv 4\pi\Phi_{ij}   \psi_{ij}.\label{Dint}
 \end{align} 
 
 The function $\tilde{I}(\mathbf{k})$ is of critical importance for accurate numerical calculations of the DDIs. For clarity we hereon refer to the analytic form of this function introduced in Eq.~(\ref{Iddk1})  as the bare $k$-space potential $ \tilde{I}_{\mathrm{bare}}$.
This result is singular since the $k\to0$ limit does not exist, reflecting the long-ranged anisotropic character of the interaction. As such the quadrature in (\ref{PhiDDij}) will converge slowly as the density of $k$-space points increases, or equivalently as the spatial ranges $R,Z$  are made larger (also see discussion in Refs.~\cite{Ronen2006a,Jian2014a}). This slow convergence can be understood as the  $\Phi$ having contributions from periodic copies (arising from using Fourier transforms on a finite interval) of the density distribution. These periodic copies, separated by twice the grid range,  interact with each other causing a shift in the interaction energy. The unphysical influence of these copies decays with the cube of the grid range, making it impractical to calculate accurate results using $ \tilde{I}_{\mathrm{bare}}$.

One approach to deal with this problem is to use spherical coordinates in which the Jacobian introduced by the coordinate transform removes the singularity of  $\tilde{I}_{\mathrm{bare}}$, but this requires the use of a  non-uniform Fourier transform \cite{Jian2014a,Bao2016a} in the algorithm. An alternative approach, first proposed for dipolar BECs in Ref.~\cite{Ronen2006a}  (also see \cite{Exl2016a,Mennemann2019a}), is to introduce a truncation of the DDI, i.e.~restrict the range of the DDI to the physical extent of the grids used, so that interactions between periodic copies are formally zero (also see \cite{Antoine2018a}, and related treatments for the Coulomb case \cite{Jarvis1997a,Rozzi2006a}). We choose to follow this approach here since it allows us to work with the cylindrical grid and associated transforms that we have introduced. The issue now becomes how to obtain the appropriate truncated DDI potential in $k$-space.

First let us consider a spherical truncation derived  and applied to dipolar BECs in Ref.~\cite{Ronen2006a}  (also see \cite{Wilson2009}). Here the truncated interaction in position space is
\begin{align}
{I}_{\mathrm{sph}}({\mathbf{r}})=\begin{cases}
\frac{3}{4\pi r^3} (1-3\cos^2 \theta), & r\le  r_c\\
0, & \mathrm{otherwise},
\end{cases}
\end{align}
i.e.~a cutoff radius $ r_c$ such that the DDIs do not occur between any two points separated by a distance of more than $2 r_c$. In practice choosing $ r_c=\min\{R,Z\}$ ensures that no interactions can occur between periodic copies of the density. Fortunately the Fourier transform of ${I}_{\mathrm{sph}}$ can be calculated analytically
\begin{align}
\tilde{I}_{\mathrm{sph}}(\mathbf{k})= &\left[1+3\frac{\cos(k r_c)}{(k r_c)^2}-
3\frac{\sin(k r_c)}{(k r_c)^3}\right] \tilde{I}_{\mathrm{bare}}(\mathbf{k}) , 
\end{align}
and is seen to regularize the behaviour of $\tilde{I}_{\mathrm{bare}}(\mathbf{k})$ near $k=0$. The spherically truncated potential is most useful for situations where the condensate is nearly spherical so it is natural to choose $R\approx Z$. 

For highly anisotropic cases the natural grid choice is $R\ll Z$ or $R\gg Z$ for cigar or pancake shaped condensates, respectively. Here the spherical truncation is impractical as the range of the narrow dimension would have to be extended to match the range of the other dimension, introducing many redundant grid points. 
In these situations it is natural to introduce a cylindrical truncation
\begin{align}
I_{\mathrm{cyl}}({\mathbf{r}})=\begin{cases}
\frac{3}{4\pi r^3} (1-3\cos^2 \theta), & \mathbf{r}\in \{\rho<R, |z|<Z\}\\
0, & \mathrm{otherwise}.
\end{cases}
\end{align}   
The Fourier transform of this truncated kernel, $\tilde{I}_{\mathrm{cyl}}$,  does not have an analytic result and needs to be calculated numerically.  
We refer the interested reader to Ref.~\cite{Lu2010a} for details about the numerical calculation.

In the testing we perform we evaluate the DDI term (\ref{Dint}) using $\Phi_{ij}$ evaluated  according to (\ref{PhiDDij}) with $\tilde{I}$ set to one of  $\{\tilde{I}_{\mathrm{bare}},\,\tilde{I}_{\mathrm{sph}},\,\tilde{I}_{\mathrm{cyl}}\}$. If not specified otherwise, we use $\tilde{I}_{\mathrm{cyl}}$

\subsection{EGPE operator and energy functional}\label{Sec:EGPEE}
We have now discussed all the operators needed to evaluate the EGPE operator 
i.e.~
\begin{align}
\mathcal{L}_{ij}\left[\psi\right]=  &h_{ij}\left[\psi \right]   + {C}_{ij}\left[\psi \right]  +  {D}_{ij}\left[\psi \right] +  {Q}_{ij}\left[\psi \right], \label{numGPEop}
\end{align}
The expectation of the EGPE operator, normalized by the field norm  gives the expected value of the chemical potential. This  is evaluated numerically as
\begin{align}
\mu_{\mathrm{EGP}}\left[\psi\right]=&\frac{1}{N[\psi]} \sum _{ij}\psi_{ij}\mathcal{L}_{ij}\left[\psi\right]dV_{i},\label{muexpt}
\end{align}
where $dV_{i}=4\pi\Delta zw_{si}$ denotes the combined quadrature weights, and $N[\psi]=\sum_{ij}\psi_{ij}^2\,dV_{i}$ is the normalization. For a stationary state  solution of the EGPE $\psi$ with eigenvalue $\mu$ [see (\ref{cylGPE})] we have that $\mu_{\mathrm{EGP}}\left[\psi\right]=\mu$.

We can also evaluate the energy functional  for the system [see Eq.~(\ref{Eq3DEfunc})]
\begin{align}
E[\psi]=&\overbrace{\sum_{ij}\psi_{ij}\left(h_{ij}\left[\psi \right]+\frac{1}{2}{C}_{ij}\left[\psi \right]\right)dV_{i}}^{E_C}    + \nonumber \\ 
&\underbrace{\frac{1}{2}\sum_{ij}\,\psi_{ij} {D}_{ij}\left[\psi \right]dV_{i}}_{E_D}
+  \underbrace{\frac{2}{5}\sum_{ij}\,\psi_{ij}{Q}_{ij}\left[\psi \right]dV_{i}}_{E_Q}, \label{EGPnum}
\end{align}
which will be a local minimum for stationary solutions of the EGPE. Here we have introduced the labels $E_C$, $E_D$ and $E_Q$ to refer to the standard GPE energy functional (including single particle and contact interactions), the dipole interaction energy and the quantum fluctuation energy, respectively.
 
\subsection{Gaussian variational solution}
It is useful to have a simple variational solution as an initial guess for the numerically calculated stationary state solutions and to validate the accuracy of our numerical methods.
Here we consider a Gaussian state with angular circulation of $s$
\begin{align}
\Psi_\text{G}(\mathbf{r})=\psi_\text{G}(\rho,z)e^{is\phi},\label{PsiVar}
\end{align}
where the cylindrical amplitude is 
\begin{align}
\psi_\text{G}(\rho,z)=\sqrt{\frac{8N}{s!\pi^{3/2}\sigma_\rho^2\sigma_z}}\left(\frac{2\rho}{\sigma_\rho}\right)^se^{-2(\rho^2/\sigma_\rho^2+z^2/\sigma_z^2)},\label{psiG}
\end{align}
with  width parameters $\{\sigma_\rho,\sigma_z\}$.

We can use this state to  analytically evaluate the separate energy terms $E_C$, $E_D$, and $E_Q$, as identified in Eq.~(\ref{EGPnum}). 
We obtain (also see \cite{Baillie2016b,Cidrim2018a})
 \begin{align}
E_{C}[\Psi_\text{G}]=&N\left(\frac{2+2s}{\sigma_\rho^2}+\frac{1}{\sigma_z^2}+\bar{\omega}_\rho^2\frac{1+s}{8}\sigma_\rho^2+\bar{\omega}_z^2\frac{\sigma_z^2}{16}\right) \nonumber \\ &+N^2\left(\frac{2}{\pi}\right)^{\frac{3}{2}}\frac{c_sg_s}{2\sigma_\rho^2\sigma_z},\label{varEC}\\ 
E_{D}[\Psi_\text{G}]=&-N^2\left(\frac{2}{\pi}\right)^{\!\frac{3}{2}} \frac{c_sg_{dd}}{2\sigma_\rho^2\sigma_z}\sum_{n=0}^{2s}d_n^sf^{(n)}(\tfrac{\sigma_\rho}{\sigma_z}),\label{varED}\\
E_{Q}[\Psi_\text{G}]=&N^{\frac{5}{2}}\frac{2^{7+5s/2}\Gamma(1+5s/2)\gamma_{\mathrm{QF}}}{(5^{1+s}s!)^{5/2}\pi^{9/4}\sigma_\rho^3\sigma_z^{3/2}},\label{varEQ}
\end{align}
where $c_s=(2s)!/4^s(s!)^2$, 
\begin{align}
f(x)=\frac{1+2x^2}{1-x^2}-\frac{3x^2\mathrm{arctanh}\sqrt{1-x^2}}{(1-x^2)^{3/2}},
\end{align}
$f^{(n)}$ denotes the $n$-th derivative of $f$ with respect to $x$, 
and 
\begin{align}
d_n^s=\begin{cases}
1, \quad& s=0,\\
1,\frac{3}{8},\frac{1}{8},&s=1,\\
1,\frac{67}{128},\frac{119}{384},\frac{3}{64},\frac{1}{384}, & s=2.
\end{cases}
\end{align}
We have also used $g_s$ and $g_{dd}$ in place of their dimensionless values introduced earlier to make it easier to transform these results to other choices of units.

The ansatz (\ref{PsiVar}) can be used to provide a variational solution to the GPE. This involves minimizing the nonlinear function
\begin{align}
E_\text{G}(\sigma_\rho,\sigma_z)=E_C[\Psi_\text{G}]+E_D[\Psi_\text{G}]+E_Q[\Psi_\text{G}], \label{EvarG}
\end{align}
to determine $\{\sigma_\rho,\sigma_z\}$.

 \begin{table*}  
  \begin{ruledtabular}
  \begin{tabular}{c|c|ccc|cccc}  
         case & $s$ & $(\sigma_\rho,\sigma_z)$&$(N_\rho,N_z) $&$(R,Z)$&$E_D\,\,(\text{bare})$&$E_D\,\,(\text{sph})$&$E_D\,\,(\text{cyl})$&$E_Q$\\
	& & \multicolumn{3}{c|}{} & \multicolumn{4}{c}{ $\log_{10}$ absolute relative error.} \\
 	\hline 
(a) & 0 & (1,10) & (25,25) & (3,30)  & -1.8 & -1.3 & -9.3 & -15 \\
(b) & 0 & (1,10) & (25,25) & (4,40)  & -2.0 & -1.7 & -15 & -15 \\
(c) & 0 & (1,10) & (250,25) & (40,40)  & -5.1 & -15 & -13 & -15 \\
(d) & 0 & (2,1) & (50,25) & (12,4)  & -1.7 & -4.8 & -15 & -15 \\
(e) & 1 & (2,1) & (50,25) & (12,4)  & -1.5 & -3.3 & -15 & -5.5 \\
(f) & 1 & (1,10) & (250,25) & (40,40)  & -4.8 & -15 & -13 & -4.4 \\
(g) & 1 & (1,10) & (256,25) & (6,40)  & -2.0 & -2.0 & -15 & -10 \\
(h) & 1 & (1,10) & (600,25) & (6,40)  & -2.0 & -2.0 & -15 & -13 \\
(i) & 1 & (1,10) & (40,25) & (6,40)  & -2.0 & -2.0 & -15 & -4.8 \\
(j) & 2 & (1,10) & (40,25) & (6,40)  & -1.9 & -1.8 & -15 & -13 \\
\hline
        \end{tabular}
  \end{ruledtabular}
      \caption{The $\log_{10}$ of the absolute relative errors of the numerically calculated values of $E_D$ and $E_Q$ compared to the analytic results [Eqs.~(\ref{varED}) and (\ref{varEQ})] for the variational Gaussian state. For all cases considered here the $E_C$ term  has an absolute relative error below $10^{-13}$ and is not given.  For the dipole energy we give results for bare, spherically-truncated and cylindrically-truncated potentials.}\label{GaussianTestTable}
     \end{table*}

\subsection{Gaussian test of numerical representation}
 \label{Sec:GaussianTest}

We use the  analytic results (\ref{varEC})-(\ref{varEQ}) to benchmark the accuracy of our numerical evaluation of the various terms appearing in the GPE. To do this we sample the variational Gaussian solution on a cylindrical grid and numerically evaluate the terms corresponding to the individual energy contributions as specified in Eq.~(\ref{EGPnum}), and compute the absolute value of the relative error with respect to the analytic results. 
  
We show the  results in Table \ref{GaussianTestTable} for various values of $s$, different grid choices and methods for evaluating the DDI term.  For all our grid choices the single particle and contact interaction term ($E_C$) has an absolute relative error of less than $10^{-13}$, so we do not list it in the table. Instead we focus on the DDI and quantum fluctuation terms that tend to have larger errors.

For the DDI term we present results for the bare, spherically truncated, and  cylindrically truncated interactions   (see Sec.~\ref{DDIterm}). Evaluating $E_D$ using the bare interaction is always inaccurate, and converges slowly to the exact result as the grid range increases [cf.~cases (a) and (b)]. The spherically cutoff interaction works well when the grid has a similar radial and axial range  [cf.~cases (b) and (c)], and is thus most useful for states where the density distribution has a similar radial and axial extent. 
The cylindrically cutoff interaction is seen to work well in all cases including highly anisotropic situations.
 
 The accuracy of the quantum fluctuation term is reduced in the $s=1$ case relative to the $s=0$ and $s=2$ cases for similar numbers of points. This finding is consistent with the analysis presented by Ogata \cite{Ogata2005a} on the accuracy of numerical Bessel quadrature (\ref{Quadr}). Ogata shows that $q$-order Bessel quadrature converges exponentially for an integrand of the form $|x|^{2q+1}h(x)dx$,  if $h(x)$ is analytic on the real axis $(-\infty,\infty)$. For our Gaussian ansatz (\ref{psiG}) the quantum fluctuation term (\ref{Q}) including the Jacobian (and neglecting $z$-coordinates) is of the form $\rho^{5s+1}g(\rho)$, where $g(\rho)$ represents the Gaussian part.
 Casting this integrand in the form Ogata analyses and taking $s=q$ we have $|\rho|^{2s+1}[|\rho^{3s}|g(\rho)]$. The term is square brackets is only analytic for $s$ even. We note that by choosing $s\ne q$  (e.g.~taking $q=1/2$) this integration can be performed more accurately, although we do not pursue this further here.

\section{Gradient flow solution of the GPE}\label{SEC:Gflow}
 
Here we present a simple gradient flow solver based on our cylindrical discretization. This is an energy minimising scheme for finding ground states\footnote{Because we constrain our EGPE to particular $s$-angular momentum spaces,  a vortex can be regarded as a ground state of that space even though it will not be the ground state of the full 3D problem.}. The gradient flow involves solving the time-dependent GPE in imaginary time, i.e.~solving the flow $\dot{\psi}=-\mathcal{L}[\psi]$. However, normalization of the field tends to decrease under this evolution, so it is necessary to renormalize during the evolution. We follow Ref.~\cite{Bao2006a} (also see \cite{Bao2010a}) and discretize the evolution using a backwards-forwards Euler scheme. Here time is advanced in time steps $\Delta t_n$, as $t_{n+1}= \Delta t_{n+1}+t_{n}$, with $n=0,1,2,\ldots$ and $t_0=0$. During such a step the updated wavefunction $\psi^+$ is obtained from the current wavefunction $\psi(t_n)$ according to 
\begin{align}
\frac{\psi^+-\psi(t_{n})}{\Delta t_{n+1}}&=\frac{1}{2}\left(D_s+\frac{\partial^2}{\partial z^2}\right)\psi^+-V_{\mathrm{eff}}[\psi(t_{n})]\psi(t_n),\label{GFGPE}\\
\psi(t_{n+1})&=\frac{\sqrt{N}\psi^+}{\sqrt{\int |\psi^+|^2d\mathbf{r}}},\label{renorm}
\end{align}
where (\ref{renorm}) is the renormalization (projection), and we have introduced
\begin{align}
    V_{\mathrm{eff}}[\psi(t_n)]\equiv&V_{\mathrm{tr}} + 4\pi\left\{\epsilon_{dd}^{-1}\psi(t_n)^2+\Phi [\psi(t_n)]\right\}  \nonumber\\
&+\gammaQF|\psi(t_n)|^3 -\mu_{\mathrm{EGP}}[\psi(t_n)].
\end{align}
Notice that the term involving the kinetic energy operator is implemented with a backwards-Euler step, giving us good stability for dense grids (where the kinetic energy operator is large).
By subtracting $\mu_{\mathrm{EGP}}$  in the $  V_{\mathrm{eff}}$ term,    we ensure that to $O(\Delta t^2)$ the field normalization is constant under the gradient flow (e.g.~see \cite{Antoine2017a}), which improves the performance of the algorithm.  Hereon we suppress explicit notation of the position indices of the wavefunction for notational brevity, however terms appearing are to be evaluated as described in Secs.~\ref{Sec:EGPEOps} and \ref{Sec:EGPEE}.

The semi-implicit equation (\ref{GFGPE}) is formally solved by inverting the spatial differential operator. This can be done  using Fourier transformation $\mathcal{F}$, yielding an explicit expression for $\psi^+$:
\begin{align}
\psi^+=\mathcal{F}^{-1}\left\{\frac{\mathcal{F}\left\{\psi(t_{n}) -\Delta t_{n+1}V_{\mathrm{eff}}[\psi(t_{n})]\psi(t_{n})\right\}}{1+\frac{1}{2}(k_\rho^2+ k_z^2)\Delta t_{n+1}}\right\}.
\end{align}
This can be efficiently implemented numerically using the operators and transforms we introduced earlier\footnote{Note that $\tilde{\psi}_{ij}=\mathcal{F}\{\psi\}$ can be evaluated as $\tilde{\psi}_{ij}=\sum_{kl}\mathcal{H}_{ik}\psi_{kl}\Lambda_{jl}$, and similarly we can define the inverse transform.}.  
 
 Using a forwards-Euler approach for the potential and nonlinear terms in Eq.~(\ref{GFGPE}) has the advantage that these terms are explicit, however care needs to be taken with the time-step to ensure that the algorithm is stable.  In practice we choose the time step according to 
\begin{align}
\Delta t_{n+1}&=\frac{a_{\Delta t}}{\norm{|V_\text{tr}| + 4\pi\epsilon_{dd}^{-1}\psi^2 + 4\pi|\Phi[\psi]| + \gamma_\text{QF}|\psi|^3}_\infty}\label{eq:maxtimeest}
\end{align}
where $||\,||_\infty$ denotes the maximum (over all spatial points) of the absolute value of the argument evaluated with $\psi(t_n)$ and $a_{\Delta t}\lesssim1$ is a constant. For the results we present here we generally take $a_{\Delta t}=0.5$\footnote{We typically find that for $a_{\Delta t}\gtrsim 1$ the gradient flow becomes unstable.}.

We are interested in obtaining the lowest energy stationary state subject to the imposed angular momentum $s$.
We can quantify the backwards error through the residual $r\equiv \mathcal{L}[\psi]-\mu\psi$. In particular we  use the measure
\begin{align}
r_\infty\equiv\frac{1}{\sqrt{N}}\norm{\mathcal{L}[\psi]-\mu\psi}_\infty,\label{rinfty}
\end{align} 
where the $N^{-1/2}$ factor is to make the measure independent of normalization choice\footnote{E.g., choosing to unit normalize the wavefunction and include the $N$ factors in the interaction parameters leaves $r_\infty$ unchanged.}.  We terminate the gradient flow once $r_\infty$ decreases below a desired value (typically $10^{-15}$). We mention that $r_\infty$  has to be used with care. First,  it depends on choice of units\footnote{E.g.~$a_{dd}\approx6.9\times10^{-9}\,$m (for Dy), whereas another popular choice of units is harmonic oscillator length, with typical values of the order of $x_\text{ho}\approx10^{-6}\,$m. Transforming a solution from $a_{dd}$-units to $x_\text{ho}$-units would increases $r_\infty$ by a factor of about $4\times10^7$. }, scaling as $x_0^{-7/2}$.  Second, because $\mu$ can be negative (i.e.~self-bound), or even zero, the sensitivity of this measure is also dependent on the case under consideration.  In practice we can evaluate (\ref{rinfty}) for a particular solution $\psi$ by applying the EGPE operator (\ref{numGPEop})  and using (\ref{muexpt}) to obtain $\mu$. Alternatively, we can approximately evaluate this as
\begin{align}
r_\infty\approx\frac{1}{\sqrt{N}}\norm{\frac{\psi(t_n)-\psi(t_{n-1})}{\Delta t_n}}_\infty,\label{rinfty2}
\end{align} 
[see Eq.~(\ref{GFGPE})].

As a second measure of solution quality we have developed a virial theorem for the EGPE.  The virial relation is $\Lambda_V=0$, where
\begin{align}
\Lambda_V=E_\text{kin}-E_\text{tr}+\frac{3}{2}E_\text{int}+\frac{9}{4}E_\text{QF},\label{EqLambdaV}
\end{align}
with the terms being the components of energy [see Eq.~(\ref{Eq3DEfunc})], and where we have assumed that any trap is harmonic in form. We have obtained this virial theorem by considering how the energy functional transforms under a scaling of coordinates (e.g.~see \cite{Dalfovo1999}).  The terms in Eq.~(\ref{EqLambdaV}) can be evaluated using the techniques described in Secs.~\ref{Sec:EGPEOps} and \ref{Sec:EGPEE}. In general we find that $|\Lambda_V|$ is a useful quantity to assess our numerical solutions, as it is sensitive to the residual $r_\infty$  and the quality of the spatial discretisation.

 \begin{table*}  
  \begin{ruledtabular} 
	 	\begin{tabular}{cccc|lll}
		& $s$ &   ($N_\rho,N_z)$   & $(R,Z) $ & \multicolumn{1}{c}{$E $}  & \multicolumn{1}{c}{$N\mu$}  & \multicolumn{1}{c}{$|\Lambda_V|$} \\ 
\hline  
(a) & 0 & (50,\,120) & (250,\,1200)  & -14.268417108 & -19.347035858 & $4.5\times10^{-3}$   \\
(b) & 0 & (80,\,160) & (400,\,1600)  & -14.268780733 & -19.351349996 & $2.3\times10^{-8}$   \\
(c) & 0 & (500,\,1000) & (500,\,2000)  & -14.268780734 & -19.351350017 & $1.2\times10^{-10}$   \\
\hline
(d) & 1 & (50,\,120) & (250,\,1200)  & -2.8144049587 & -9.5969313711 & $1.8\times10^{-6}$   \\
(e) & 1 & (80,\,160) & (400,\,1600)  & -2.8144048901 & -9.5969333804 & $3.6\times10^{-7}$   \\
(f) & 1 & (200,\,480) & (400,\,1600)  & -2.8144047844 & -9.5969330684 & $1.2\times10^{-9}$   \\
(g) & 1 & (500,\,1000) & (500,\,2000)  & -2.8144047842 & -9.5969330677 & $3.4\times10^{-10}$   \\
  \end{tabular}
  \end{ruledtabular} 
\caption{Energy and chemical potential values for  $\epsilon_{dd}^{-1}=0.5$ droplets with $N=10^4$ and  (a)-(c) $s=0$ or (d)-(g) $s=1$ on various choices of numerical grid. All results are calculated with a cylindrically cutoff DDI, and the gradient flow is terminated when $r_\infty$ decreases below $3\times10^{-17}$.}
\label{Tab:GFfors0s1}  
\end{table*}

\begin{figure}[htbp] 
  \centering
   \includegraphics[width=3.2in]{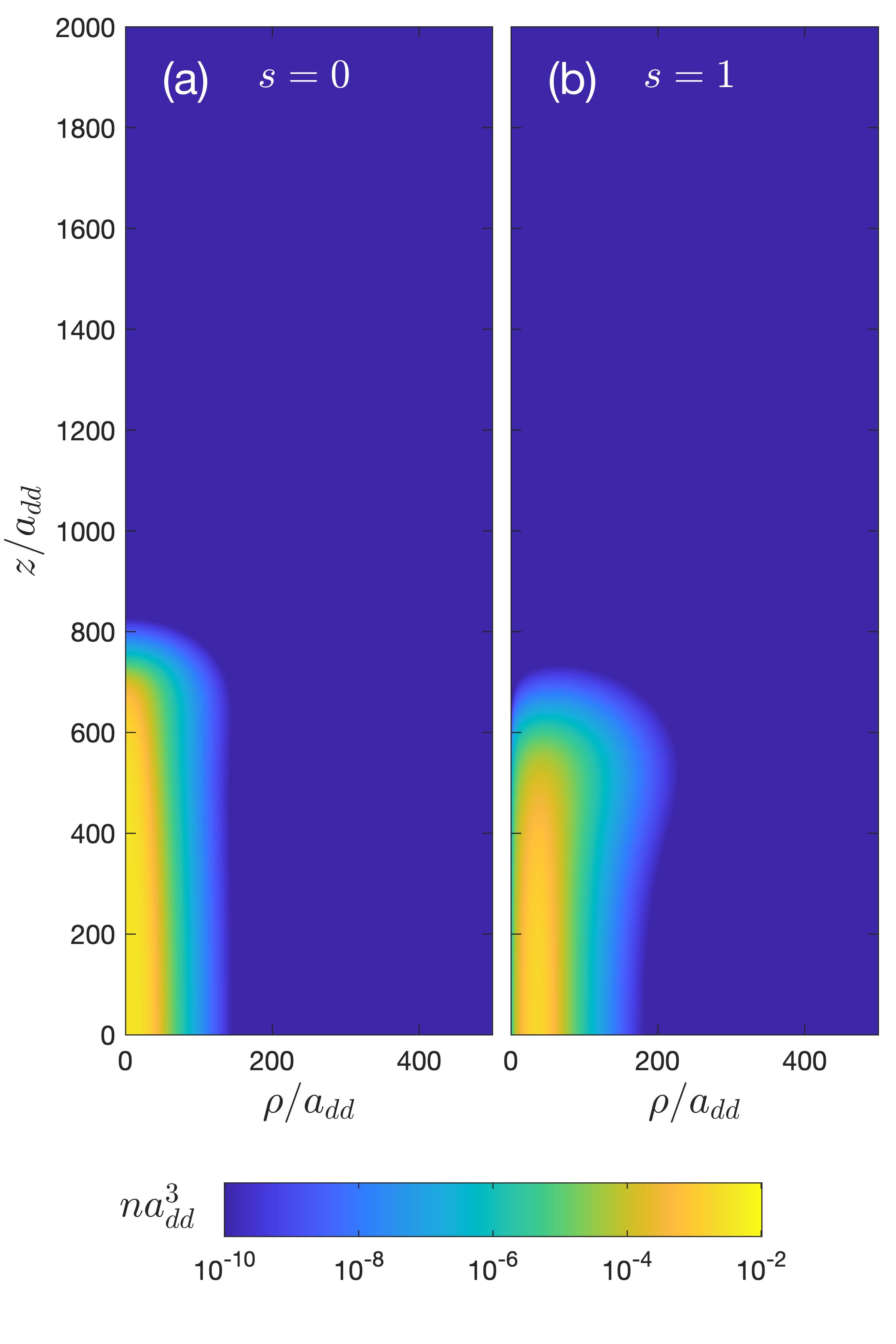}
  \caption{Density profiles of  (a) $s=0$ and (b) $s=1$ self-bound stationary states for a system with $\epsilon_{dd}^{-1}=0.5$ and $N=10^4$ atoms. (c)-(e) Gradient flow evolution for case (b) of Table \ref{Tab:GFfors0s1} using $a_{\Delta t}=0.5$, showing (c) the error $r_\infty$, (d) time step $\Delta t_n$,  (e) energy error, and (f) the absolute virial expectation error. In the energy error $E_\text{ref}$ is a reference energy calculated from a state using a larger more dense grid.}
     \label{fig:states}
\end{figure}

\section{Results}\label{SEC:Results}
In Table \ref{Tab:GFfors0s1} we present results for the energy, chemical potential and $|\Lambda_V|$ of stationary self-bound states with $s=0$ and $s=1$ obtained by the gradient flow method.  The density profiles of the states are shown in Figs.~\ref{fig:states}(a) and (b). We also show the gradient flow evolution for case (b) of Table \ref{Tab:GFfors0s1} in Figs.~\ref{fig:states}(c)-(f).  We use the variational solution $\psi_\text{G}$ as the initial condition for the  gradient flow  and the time step $\Delta t_n$ quickly settles to a value of $\Delta t_n\approx6.4$ (for $a_{\Delta t}=0.5$) during the flow. The flow terminates at $r_\infty=5\times10^{-17}$ after 5591 steps, taking about 15 seconds to execute on a workstation computer. For this case the flow is stable for $a_{\Delta t}\lesssim0.84$, and takes a proportionately shorter amount of time to execute with a larger value of $a_{\Delta t}$, but becomes unstable and does not converge when $a_{\Delta t}\gtrsim0.84$.  

The results in Table \ref{Tab:GFfors0s1} reveal how sensitive the energy and chemical potential are to the choice of grid [i.e.~range $(R,Z)$ and number of points $(N_\rho,N_z)$], showing that accurate results (greater than 9 significant figures) can be obtained when the grid ranges are approximately twice the spatial extent of this droplet and for sufficiently many points. This grid range dependence arises because the DDI term involves a convolution (e.g.~see discussion in Ref.~\cite{Greengard2018a}). The $s=1$ case  typically requires more spatial points to have a similar level of accuracy as the $s=0$ case. We expect that this arises from the poorer performance of the Bessel quadrature for the quantum fluctuation term when $s=1$, as noted in Sec.~\ref{Sec:GaussianTest}. 
The results in the table also show that the absolute virial expectation $|\Lambda_V|$  is qualitative similar to the magnitude to the energy error (i.e.~number of converged digits in $E$), suggesting that $|\Lambda_V|$ provides a useful additional method to characterize the solution accuracy.

 We also consider the calculation of a phase diagram to predict where $s=0$ and $s=1$ self-bound droplets have a negative energy. This requirement ensures that the droplet is energetically stable against evaporating into the trivial   $E=0$ state where the atoms are dispersed over all space. This condition is adequate to ensure that $s=0$ droplets are ground states that are dynamically stable. For the $s=1$ case this requirement does not ensure dynamic stability as the droplet can decay by fission. We discuss this aspect later.
 
\begin{figure}[htbp] 
  \centering
   \includegraphics[width=3.4in]{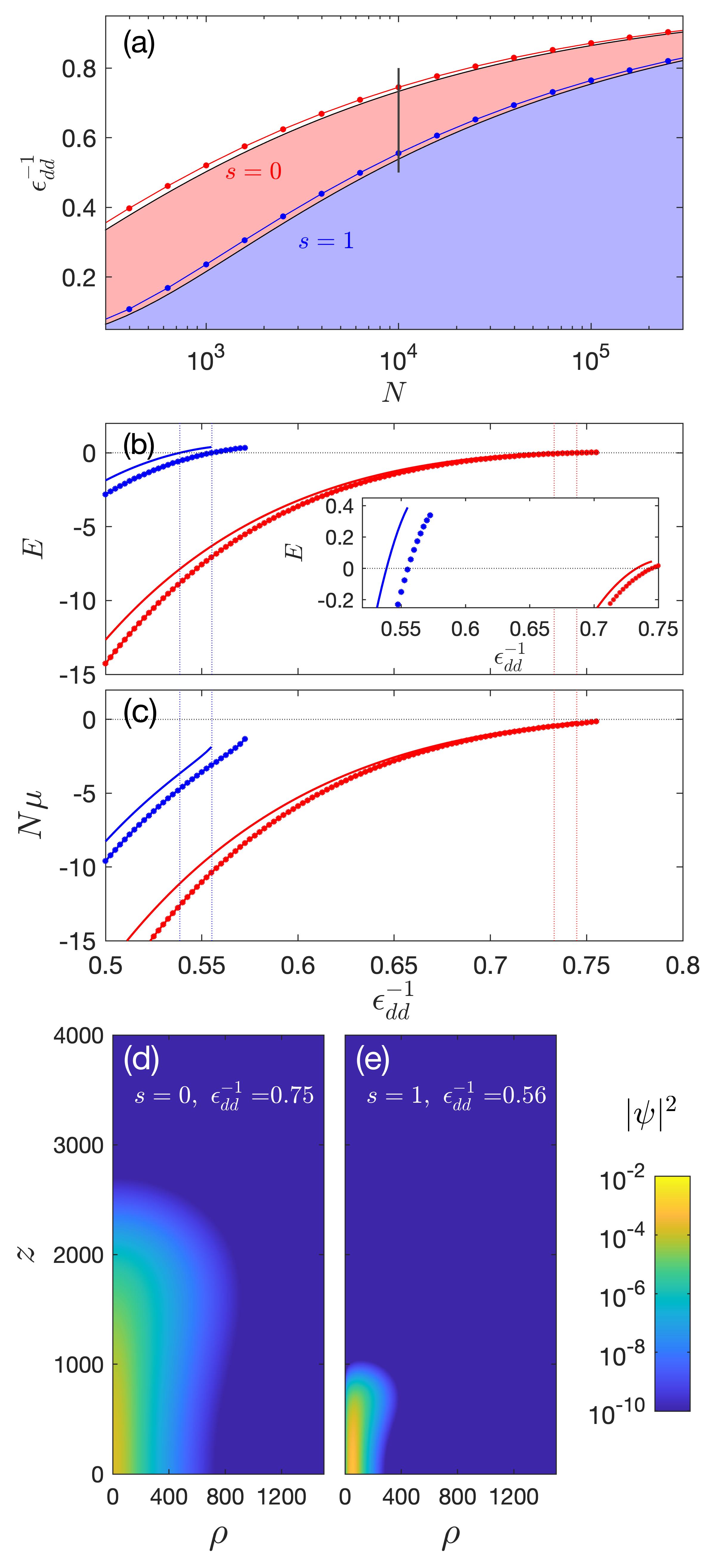}
   \caption{ (a) Phase diagram of self-bound $s=0$ and $s=1$ droplet solutions. Shaded regions bordered by black lines mark where $E_\text{G}<0$. Lines with circle markers show where the EGPE solution has $E=0$. (b) The energy and (c) $N\mu$ versus $\epsilon_{dd}^{-1}$ for $N=10^4$ for the variational (lines) and GPE (lines with small circles) results for $s=0$ (red) and $s=1$ (blue). The dotted lines indicate where $E=0$. The inset shows the data near $E=0$. Subplots (d) and (e) show examples of metastable states with positive energies. (d) $s=0$ droplet with $\epsilon_{dd}^{-1}=0.75$, $E=1.7634\times10^{-2}$, $\mu=-2.1742\times10^{-5}$, and (e) $s=1$, $\epsilon_{dd}^{-1}=0.56$, $E=1.1746\times10^{-1}$, $\mu=-2.6444\times10^{-4}$.}
   \label{fig:phasediag}
\end{figure}

Our results for the regions where the droplet states have negative energy are shown in Fig.~\ref{fig:phasediag}(a). Note by using coordinates of $N$ and $\epsilon_{dd}^{-1}$ this phase diagram has no remaining dependence on other parameters, and is in this sense universal.  In general the $s=0$ and $s=1$ regions have similar shapes, although the $s=0$ region is larger (extends to higher $\epsilon_{dd}^{-1}$). This is because the $s=0$ droplet has a considerably lower energy for the same parameters [e.g.~see Fig.~\ref{fig:phasediag}(b) and Table \ref{Tab:GFfors0s1}]. This difference is from the large energy cost associated with hosting a vortex in the droplet. This arises from the kinetic energy of the vortex, but also because the $s=1$ droplet is wider than the $s=0$ droplet [cf.~Figs.~\ref{fig:states}(a) and (b)], making the DDI energy less negative.

 The results in Fig.~\ref{fig:phasediag} also compare the stationary EGPE solutions and variational Gaussian approach.
For example, the markers in Fig.~\ref{fig:phasediag}(a) show the boundary to the negative energy region determined by the EGPE solutions, while the boundary of the shaded region is determined variationally.  The EGPE results are obtained using the gradient flow method to solve for a state at an initial (low) value of $\epsilon_{dd}^{-1}$ stating from a variational Gaussian solution. The resulting EGPE solution for that case is then used as the starting point for the gradient flow at a slightly higher value of $\epsilon_{dd}^{-1}$, and so on. This process is stopped once the localized state disperses (unbinds and spreads out over the range of the grid). In Fig.~\ref{fig:phasediag}(b) we show the energy of a sequence of states obtained this way for a particular case of $N=10^4$. We also indicate the point where $E=0$, used to identify the boundary.  The variational results are located as an energy minimum of the  function given in Eq.~(\ref{EvarG}). We follow this minimum as $\epsilon_{dd}^{-1}$ increases, noting that eventually it becomes a local minimum (when $E>0$) and then changes to a saddle point (causing the branch to terminate). 
We also mention that while the energy is positive near the end of the branches, the chemical potential remains negative [Fig.~\ref{fig:phasediag}(c)]. 

The phase diagram in Fig.~\ref{fig:phasediag}(a) collates the results of many analyses of the type in Fig.~\ref{fig:phasediag}(b)  for different atom numbers $N$.   In general the $E=0$ boundary predicted by the variational Gaussian theory is close to that obtained from the EGPE solution. This is because the energies predicted by the two approaches are similar near the transition point. However, we note that the energy of the EGPE state tends to be significantly lower than the variational state for smaller values of $\epsilon_{dd}^{-1}$ where the droplets are more deeply bound. In this regime [e.g.~ $s=0$ state in Fig.~\ref{fig:states}(a)] the droplet density profile has a relative flat-top axial density distribution, which is not well-captured by a Gaussian.

In Figs.~\ref{fig:phasediag}(d) and (e) we show examples of the droplet states for cases with slightly positive energy. We can compare these states to the more deeply bound states (differing only in the values of $\epsilon_{dd}^{-1}$) from Fig.~\ref{fig:states}. This comparison, particularly for the $s=0$ case, emphasizes how much the droplet size can change with $\epsilon_{dd}^{-1}$. For example, over the range of $\epsilon_{dd}^{-1}$ considered in Fig.~\ref{fig:phasediag}(b) the axial length of the $s=0$ state changes by approximately a factor of 5.  This necessitates careful choice of numerical girds to ensure the states are calculated accurately as $\epsilon_{dd}^{-1}$ changes.

\begin{figure}[htbp] 
  \centering
   \includegraphics[width=3.4in]{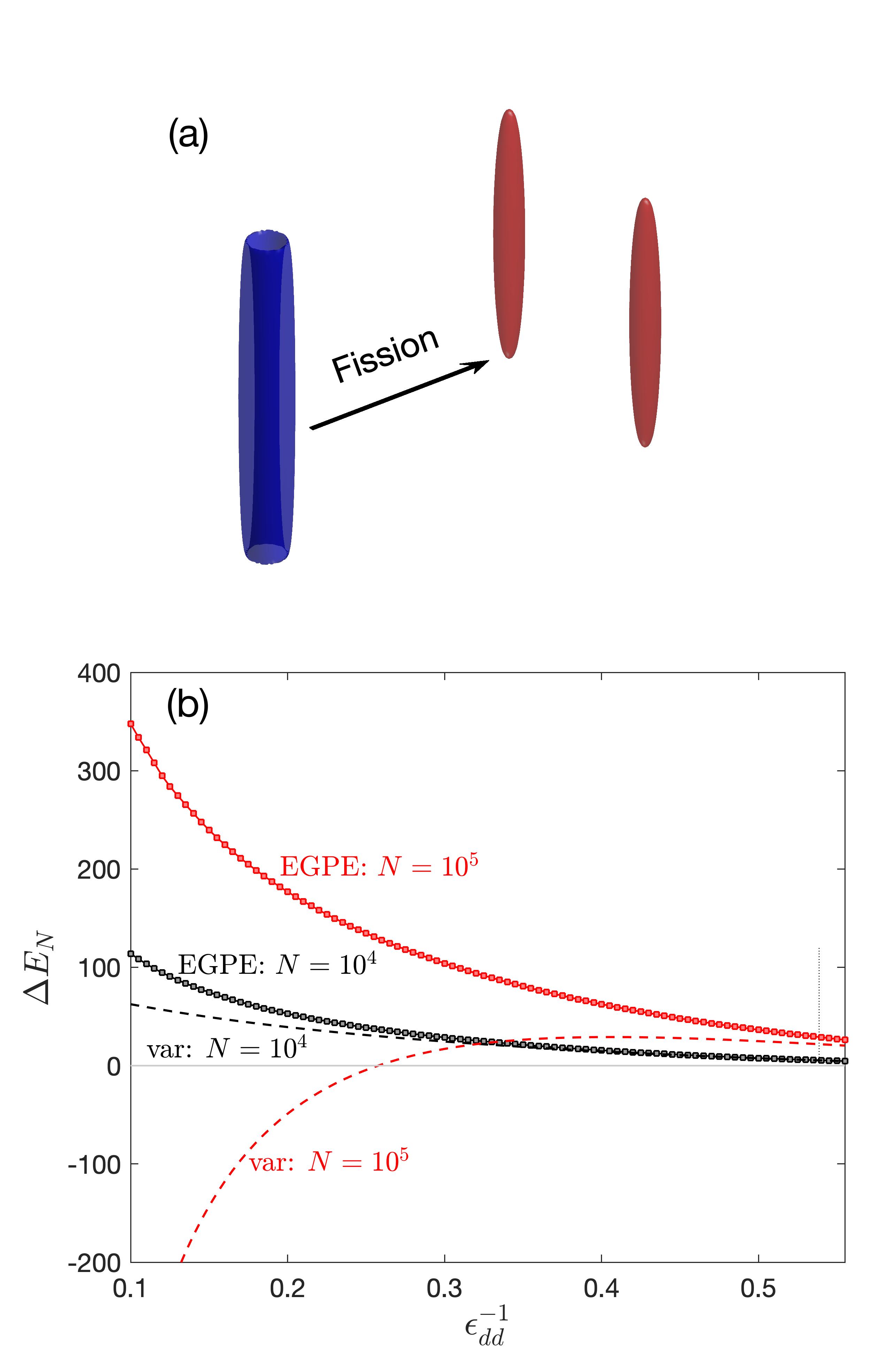}
   \caption{Vortex droplet fission. (a) A schematic of the fission process where a $s=1$ vortex droplet (blue) decays into two $s=0$ droplets (red). (b) Fission energy $\Delta E_N$ as a function of $\epsilon_{dd}^{-1}$ for $N=10^4$ (black lines/markers) and $N=10^5$  (red lines/markers). The results are calculated from the variational (lines) and EGPE (markers) theories.  The vertical dotted line indicates where the $N=10^4$ $s=1$ droplet has zero-energy.
   \label{fig:DE}}
\end{figure}

We can assess the stability of the $s=1$ vortex droplet to undergoing fission, whereby it  splits into two $s=0$ droplets [see Fig.~\ref{fig:DE}(a)]. This scenario has been observed in dynamical simulations of the $s=1$ vortex droplet \cite{Cidrim2018a}. We can define a fission energy as
\begin{align}
\Delta E_N= E^{s=1}_N-2E^{s=0}_{\frac{N}{2}},
\end{align}
where $E^{s=1}_N$ is the energy of a $s=1$ vortex droplet with $N$ atoms, and $E^{s=0}_{\frac{N}{2}}$ is the energy of a $s=0$ droplet with $N/2$ atoms. Thus $\Delta E_N$ represents the excess energy of the vortex droplet compared to two independent  (i.e.~well-separated) $s=0$ droplets. When this energy is positive, then the vortex is potentially unstable to fission. The angular momentum of the vortex droplet would need to be carried by the motion of the separating $s=0$ droplets, meaning that kinetic energy will also be important in determining if fission can occur.   Thus $\Delta E_N>0$ is a necessary but not sufficient condition for fission.

In Fig.~\ref{fig:DE}(b) we show some results for the fission energy computed using the EGPE and variational solutions. For the results with $N=10^4$ both approaches predict that $\Delta E_N>0$, and thus the vortex droplet is potentially unstable to fission. However, for the larger atom number of $N=10^5$ the two predictions are different: the EGPE results predict that  $\Delta E_N>0$ at all values of $\epsilon_{dd}^{-1}$ considered, while the variational results suggest stability (i.e.~$\Delta E_N<0$) for  $\epsilon_{dd}^{-1}\lesssim0.25$. Here the variational approximation is failing because the droplets are not well described by the Gaussian ansatz. We note that this problem was originally considered in a preprint by Cidrim \textit{et al.}~\cite{Cidrim2017a}, who presented EGPE results that appeared to support the variational prediction that $\Delta E_N<0$ for large $N$ and small $\epsilon_{dd}^{-1}$. Because $\Delta E_N$ is determined by a difference in two calculations, it can be quite sensitive to the accuracy of the EGPE calculations. This example emphasizes the need for high accuracy calculations of dipolar droplets.   

\section{Conclusions}\label{SEC:Conclusions}
 In this paper we have presented a method to accurately solve for dipolar quantum droplets in a cylindrical geometry allowing for the inclusion of angular momentum.  This work builds on the discretization introduced by Ronen \textit{et al.}~\cite{Ronen2006a}, extending it to include angular momentum in the stationary state, a cylindrical cutoff (truncation) of the DDI kernel, and the quantum fluctuation term. Using a simple gradient flow technique we demonstrate that this approach is able to obtain accurate results  for the dense, highly elongated (filament shaped) quantum droplets that form in dipolar BECs in the regime where the DDIs dominate the contact interactions. We also show that without a careful treatment of the DDI term in this regime it may be difficult to obtain the droplet energy to better than $\sim$10\% accuracy. Such errors would make calculations such as the fission energy (which depends in the difference of energies between two states) difficult to compute.

 We have also presented  benchmark energy and chemical potential calculations for self-bound droplets. There are very few such benchmark results in the literature and we expect these will be important for comparisons with different approaches that may be developed in the future. We present a generalization of the virial theorem for dipolar EGPE and find that this can be used to test the solution accuracy. As an application of our method we have also presented a phase diagram for the energetic stability of self-bound $s=0$ and $s=1$ droplets, and considered the stability of $s=1$ droplets against fission.
  
There are many avenues for future development of this work.
We note that the EGPE solutions we have presented were obtained using a simple but efficient backwards-forwards Euler gradient flow method. It would be of interest to consider other more efficient solvers such as conjugate gradient solvers (e.g.~see \cite{Ronen2006a,Antoine2017a,Antoine2018a}) or a fully implicit backwards-Euler method.  Such solvers will be useful in situations where efficiency becomes more important, such as fully three-dimensional (3D) cases that cannot be reduced to cylindrical symmetry. Typically, non-cylindrically symmetric ground states occur when the confinement is not symmetric about the dipole polarization axis (e.g.~\cite{Blakie2020a}), or when the system favours the formation of a droplet array (including supersolid) \cite{Baillie2018a,Wenzel2017a}. This latter case is of significant interest in the community due to the recent observation of supersolid states \cite{Tanzi2019a,Chomaz2019a,Tanzi2019b,Guo2019a,Natale2019a,Hertkorn2019a}. Another area requiring attention in the fully 3D case is an efficient and accurate truncation scheme for the DDI kernel, which is a critical tool to enable an efficient grid choice to represent the state.

\section{Acknowledgements}
We acknowledge support from the Marsden Fund of the Royal Society of New Zealand, and valuable discussions with Y.~Cai and W.~Bao.
   
%

 \end{document}